\documentclass[11pt,a4paper]{article}
\usepackage{amsmath}
\usepackage[mathcal]{eucal}
\usepackage{latexsym}
\usepackage{amssymb}
\begin{titlepage}
\title{Quantum algebra of the Hamiltonian constraint in reduced 4-dimensional gravity}
\author{Eyo Eyo Ita III}
\medskip
\input amssym.def
\input amssym.tex

\def \in{\indent}

\begin{document}
\maketitle
\bigskip
\centerline{Department of Applied Mathematics and Theoretical Physics} 
\smallskip
\centerline{Centre for Mathematical Sciences, University of Cambridge, Wilberforce Road}
\smallskip
\centerline{Cambridge CB3 0WA, United Kingdom}
\smallskip
\centerline{eei20@cam.ac.uk} 

\bigskip

\begin{abstract}
In this paper we demonstrate closure of the quantum algebra of Hamiltonian constraints in a theory directly related to a certain sector of general relativity reduced to diagonal variables.
\end{abstract}
\end{titlepage}

\section{Introduction}

In \cite{EYOITA} an action was presented which is related to a certain sector of reduced gravity, and it was demonstrated that the associated system is Dirac consistent at the classical level as a stand-alone action.  By reduced gravity, we mean that the theory has three degrees of freedom per point at the kinematic level, namely the level prior to implementation of the Hamiltonian constraint.  In this theory there are no Gauss' law and no diffeomorphism constraints.  Additionally, there are six distinct sectors, referred to as quantizable configurations $\Gamma_q$, which exhibit the same features as outlined in \cite{EYOITA}.  The results of the present paper will also apply to these configurations $\Gamma_q$.  The purpose of the present paper will be to verify closure of the algebra of constraints for the action written down in \cite{EYOITA} at the quantum level.  In this section we will introduce the action, and in section 2 we carry out the computation of the quantum constraints algebra.\par
\indent  
Let us consider a system with configuration and momentum space variables $\Gamma_{Kin}=(X,Y,T)$ and $P_{Kin}=(\Pi_1,\Pi_2,\Pi)$ defined on a 4-dimensional spacetime manifold of topology $M=\Sigma\times{R}$, where $\Sigma$ is a 3-dimensional spatial hypersurface.  The variables are in general complex, and the configuration space variables take on the ranges $-\infty<\vert{X}\vert,\vert{Y}\vert,\vert{T}\vert<\infty$.  The mass dimensions of all variables have been chosen to be 
\begin{eqnarray}
\label{JACOBI8}
[\Pi_1]=[\Pi_2]=[\Pi]=1;~~[X]=[Y]=[T]=0.
\end{eqnarray}
\noindent
These variables define the following kinematic phase space action for a totally constrained system
\begin{eqnarray}
\label{STARTINGACTIONN}
I_{Kin}=-{i \over G}\int{dt}\int_{\Sigma}d^3x\bigl(\Pi_1\dot{X}+\Pi_2\dot{Y}+\Pi\dot{T}\bigr)-iH[N].
\end{eqnarray}
\noindent
The field $N$ is an auxilliary field smearing a phase space function $H$, such that the Hamiltonian density is given by
\begin{eqnarray}
\label{STUFF}
H[N]=\int_{\Sigma}d^3xNUe^{-T/2}\Phi.
\end{eqnarray}
\noindent
The quantities in (\ref{STUFF}) are defined as follows.  First we have $\Phi$, given by
\begin{eqnarray}
\label{ALBOOT}
\Phi=\sqrt{\Pi(\Pi+\Pi_1)(\Pi+\Pi_2)}
\Bigl[\Bigl(k+e^T\Bigl({1 \over \Pi}+{1 \over {\Pi+\Pi_1}}+{1 \over {\Pi+\Pi_2}}\Bigr)\Bigr]
\end{eqnarray}
\noindent
where $k$ is a numerical constant.\footnote{For the reduced sector of gravity, we must have $k=\Lambda{a}_0^{-3}$, 
where $\Lambda$ is the cosmological constant and $a_0$ is a numerical constant of mass 
dimension $[a_0]=1$.}  There are no spatial derivatives in any of the quantities in (\ref{ALBOOT}), and all spatial derivatives in the theory (\ref{STARTINGACTIONN}) are confined to the 
quantity $U$, given by
\begin{eqnarray}
\label{YOU}
U=\Bigl[1+e^{-T}\bigl((\partial_2Z)(\partial_3X)(\partial_1Y)-(\partial_3Y)(\partial_1Z)(\partial_2X)\bigr)\nonumber\\
+e^{-2X}(\partial_1Y)(\partial_1Z)+e^{-2Y}(\partial_2Z)(\partial_2X)+e^{-2Z}(\partial_3X)(\partial_3Y)\Bigr]^{1/2}
\end{eqnarray}
\noindent
with $Z=T-X-Y$.  We have defined
\begin{eqnarray}
\label{DEEFE}
\partial_1={\partial \over {\partial{y}^1}};~~\partial_2={\partial \over {\partial{y}^2}};~~\partial_3={\partial \over {\partial{y}^3}},
\end{eqnarray}
\noindent
where $y^1$, $y^1$ and $y^3$ are dimensionless spatial coordinates in $\Sigma$.\par
\indent
The canonical structure of (\ref{STARTINGACTIONN}) yields the following fundamental Poisson brackets
\begin{eqnarray}
\label{THEFOLLOWING}
\{{X}(x,t),{\Pi}_1(y,t)\}=\{{Y}(x,t),{\Pi}_2(y,t)\}=\{{T}(x,t),{\Pi}(y,t)\}=-iG\delta^{(3)}(x,y).
\end{eqnarray}
\noindent
In this paper we will check for closure of the quantum constraints algebra of (\ref{STARTINGACTIONN}).  But prior to proceeding with the algebra, it is worthwhile to present a short background of the significance of the 
action (\ref{STARTINGACTIONN}).

\subsection{Relation to an antecedent of the CDJ action}
  
The significance of the action (\ref{STARTINGACTIONN}) is that it can be obtained from a restricted sector of an action for general relativity, which appears in \cite{SPINCON} as an intermediate step in obtaining the CDJ pure spin connection formulation for gravity,\footnote{The initial CDJ refer to Capovilla, Dell and Jacobson, who developed a nonmetric formulation for gravity written almost completely in terms of the spin connection.} as we will demonstrate.  
Consider the following change of variables of the action (\ref{STARTINGACTIONN})
\begin{eqnarray}
\label{TRANSFORMATION}
\Pi=a_0^3e^T\lambda_3;~~\Pi+\Pi_1=a_0^3e^T\lambda_1;~~\Pi+\Pi_2=a_0^3e^T\lambda_2
\end{eqnarray}
\noindent
for the momentum space variables $P_{Kin}$, and
\begin{eqnarray}
\label{TRANSFORMATION1}
X=\hbox{ln}\Bigl({{a_1} \over {a_0}}\Bigr);~~Y=\hbox{ln}\Bigl({{a_2} \over {a_0}}\Bigr);~~T=\hbox{ln}\Bigl({{a_1a_2a_3} \over {a_0^2}}\Bigr)
\end{eqnarray}
\noindent
for the configuration space variables $\Gamma_{Kin}$, where $a_0$ is a numerical constant of mass dimension $[a_0]=1$.  Let us also make the definitions
\begin{eqnarray}
\label{TRANSFORMATION2}
x^1={{y^1} \over {a_0}};~~x^2={{y^2} \over {a_0}};~~x^3={{y^3} \over {a_0}}
\end{eqnarray}
\noindent
with $y^1$, $y^2$ and $y^3$ the dimensionless spatial coordinates in $\Sigma$.  This implies that $[x^1]=[x^2]=[x^3]=-1$, namely that the coordinates $x^1$, $x^2$ and $x^3$ have dimensions of 
length.  Substitution of (\ref{TRANSFORMATION1}) and (\ref{TRANSFORMATION2}) into (\ref{YOU}) yields
\begin{eqnarray}
\label{TRANSFORMATION3}
U=(a_1a_2a_3)^{-1}\biggl[(a_1a_2a_3)^2+(\partial_2a_3)(\partial_3a_1)(\partial_1a_2)\nonumber\\
-(\partial_3a_2)(\partial_1a_3)(\partial_2a_1)+a_2a_3(\partial_1a_2)(\partial_1a_3)+a_3a_1(\partial_2a_3)(\partial_2a_1)\nonumber\\
+a_1a_2(\partial_3a_1)(\partial_3a_2)\biggr]^{1/2}=(\hbox{det}A)^{-1}(\hbox{det}B)^{1/2},
\end{eqnarray}
\noindent
from which one recognizes $U$ as the square root of the determinant of the magnetic field $B^i_a$ for a diagonal connection $A^a_i=diag(a_1,a_2,a_3)$, with the leading order term in $(\hbox{det}A)$ factored out.  In matrix form this is given by
\begin{displaymath}
a^a_i=
\left(\begin{array}{ccc}
a_1 & 0 & 0\\
0 & a_2 & 0\\
0 & 0 & a_3\\
\end{array}\right);~~
b^i_a=
\left(\begin{array}{ccc}
a_2a_3 & -\partial_3a_2 & \partial_2a_3\\
\partial_3a_1 & a_3a_1 & -\partial_1a_3\\
-\partial_2a_1 & \partial_1a_2 & a_1a_2\\
\end{array}\right)
.
\end{displaymath}
\noindent
Substitution of (\ref{TRANSFORMATION}), (\ref{TRANSFORMATION1}) and (\ref{TRANSFORMATION3}) into (\ref{STARTINGACTIONN}) yields
\begin{eqnarray}
\label{REPEAT}
I=-{i \over G}\int{dt}\int_{\Sigma}d^3x\Bigl(\lambda_1a_2a_3\dot{a}_1+\lambda_2a_3a_1\dot{a}_2+\lambda_3a_1a_2\dot{a}_3\nonumber\\
-iN(\hbox{det}b)^{1/2}\sqrt{\lambda_1\lambda_2\lambda_3}\Bigl(\Lambda+{1 \over {\lambda_1}}+{1 \over {\lambda_2}}+{1 \over {\lambda_3}}\Bigr).
\end{eqnarray}
\noindent
Let us make the following definitions for the magnetic field and the temporal component of the curvature
\begin{eqnarray}
\label{DEFINITIONSUSED}
b^i_a={1 \over 2}\epsilon^{ijk}f^a_{jk};~~f^a_{0i}=\dot{a}^a_i-D_ia^a_0,
\end{eqnarray}
\noindent
where $D_i$ is the $SO(3,C)$ covariant derivative with respect to the spatial connection $a^a_i$.  Then the integrand of the canonical one form of (\ref{REPEAT}) can be written as
\begin{eqnarray}
\label{THEINTEGRAND}
\lambda_gb^i_g\dot{a}^g_i={1 \over 2}\lambda_g\epsilon^{ijk}f^g_{jk}f^g_{0i}+\lambda_gb^i_gD_ia^g_0,
\end{eqnarray}
\noindent
where $a^g_0$ is the temporal component of the connection $a^g_{\mu}$.  Then defining $\epsilon^{ijk}=\epsilon^{0ijk}$ and using the symmetries of the 4-D epsilon symbol $\epsilon^{\mu\nu\rho\sigma}$, then (\ref{THEINTEGRAND}) is given by
\begin{eqnarray}
\label{THEINTEGRAND1}
{1 \over 8}\lambda_gf^g_{\mu\nu}f^g_{\rho\sigma}\epsilon^{\mu\nu\rho\sigma}-a^g_0b^i_gD_i\lambda_g.
\end{eqnarray}
\noindent
Using equation (\ref{THEINTEGRAND1}) to replace the canonical one form in (\ref{REPEAT}), we get the action
\begin{eqnarray}
\label{THEINTEGRAND2}
I=-{i \over G}\int_Md^4x\Bigl[{1 \over 8}\lambda_gf^g_{\mu\nu}f^g_{\rho\sigma}\epsilon^{\mu\nu\rho\sigma}\nonumber\\
-i\eta\Bigl(\Lambda+{1 \over {\lambda_1}}+{1 \over {\lambda_2}}+{1 \over {\lambda_3}}\Bigr)\Bigr]
+\int{dt}\int_{\Sigma}d^3xa^g_0b^i_gD_i\lambda_g,
\end{eqnarray}
\noindent
where $\eta=(\hbox{det}b)^{1/2}\sqrt{\lambda_2\lambda_2\lambda_3}$.  Equation (\ref{THEINTEGRAND2}) is none other than the CDJ antecedent appearing in \cite{SPINCON} with the following caveats: (i) The Gauss' law constraint is missing.  This is the last term on the right hand side of (\ref{THEINTEGRAND2}), which cancels the same quantity from the curvature squared term. (ii) Equation (\ref{THEINTEGRAND2}) is the restriction of the aforementioned action to diagonal variables.\par
\indent

\section{Quantum constraints algebra of the Hamiltonian constraint}

Upon quantization of (\ref{STARTINGACTIONN}), the dynamical variables become promoted to quantum operators satisfying equal-time commutation relations
\begin{eqnarray}
\label{THEFOLLOWINGONE}
\bigl[\hat{X}(x,t),\hat{\Pi}_1(y,t)\bigr]=\bigl[\hat{Y}(x,t),\hat{\Pi}_2(y,t)\bigr]=\bigl[\hat{T}(x,t),\hat{\Pi}(y,t)\bigr]=(\hbar{G})\delta^{(3)}(x,y),
\end{eqnarray}
with all other commutators vanishing.  The smeared Hamiltonian constraint (\ref{STUFF}) becomes promoted to a composite operator constraint
\begin{eqnarray}
\label{ALGEBRA12}
\hat{H}[N]=\int_{\Sigma}d^3xN(x)\hat{\eta}(x)\hat{\Phi}(x),
\end{eqnarray}
\noindent
where we have made the following definitions
\begin{eqnarray}
\label{ALGEBRA11}
\hat{\eta}(x)=\hat{U}(x)e^{-\hat{T}(x)/2};\nonumber\\
\hat{\Phi}=\sqrt{\hat{\Pi}(\hat{\Pi}+\hat{\Pi}_1)(\hat{\Pi}+\hat{\Pi}_2)}
\Bigl[k+\Bigl({1 \over {\hat{\Pi}}}+{1 \over {\hat{\Pi}+\hat{\Pi}_1}}+{1 \over {\hat{\Pi}+\hat{\Pi}_2}}\Bigr)e^{\hat{T}}\Bigr],
\end{eqnarray}
\noindent
with the operator ordering as indicated.  The physical states are defined as those states $\bigl\vert\psi\bigr>\in\bigl\vert\psi_{Phys}\bigr>$ such that $\hat{H}\bigl\vert\psi\bigr>=0$ with $\hat{\Phi}$ appearing to the right.  For the quantum constraints algebra to be consistent in the Dirac sense, the algebra must close with $\hat{\Phi}$ appearing on the right.\par  
\indent
References \cite{ORDER} and \cite{SCHRO6}) state that the quantization of theories containing operator products evaluated at the same point results in infinities which need to be regularized.  A possible regularization prescription is to individually smear each operator appearing in the operator product.  We will show that such regularization procedures are not necessary for the case presented in this paper, since the smearing of the constraints automatically eliminates any infinitites.  For the quantum constraints algebra we will use the following operator identity for composite operators
\begin{eqnarray}
\label{ALGEBRA10}
\bigl[\hat{A}\hat{B},\hat{C}\hat{D}\bigr]=\hat{C}[\hat{A},\hat{D}]\hat{B}+\hat{A}[\hat{B},\hat{C}]\hat{D}
+[\hat{A},\hat{C}]\hat{B}\hat{D}+\hat{C}\hat{A}[\hat{B},\hat{D}],
\end{eqnarray}
\noindent
where $\hat{A}$, $\hat{B}$, $\hat{C}$ and $\hat{D}$ are bosonic operators.  Using equation (\ref{ALGEBRA10}), the quantum constraints algebra of the Hamiltonian (\ref{ALGEBRA12}) is given by
\begin{eqnarray}
\label{ALGEBRA13}
\bigl[\hat{H}[M],\hat{H}[N]\bigr]=\int_{\Sigma}d^3x\int_{\Sigma}d^3yM(x)N(y)\bigl[\hat{\eta}(x)\hat{\Phi}(x),\hat{\eta}(y)\hat{\Phi}(y)\bigr]\nonumber\\
=\int_{\Sigma}d^3x\int_{\Sigma}d^3yM(x)N(y)\biggl[\hat{\eta}(y)[\hat{\eta}(x),\hat{\phi}(y)]\hat{\phi}(x)
+\hat{\eta}(y)[\hat{\Phi}(x),\hat{\eta}(y)]\hat{\Phi}(y)\nonumber\\
+[\hat{\eta}(x),\hat{\eta}(y)]\hat{\Phi}(x)\hat{\Phi}(y)+\hat{\eta}(x)\hat{\eta}(y)[\hat{\Phi}(x),\hat{\Phi}(y)]\biggr].
\end{eqnarray}
\noindent
We must now analyse each term appearing in (\ref{ALGEBRA13}).  The third term on the right hand side of (\ref{ALGEBRA13}) vanishes since it is a commutator purely between configuration space 
variables.  The fourth term of (\ref{ALGEBRA13}) vanishes, which can be seen as follows.  Make the following definitions
\begin{eqnarray}
\label{SHOWIT}
\hat{W}(x,y)=\hat{\eta}(x)\hat{\eta}(y);~~
\hat{\Phi}(x)\equiv\hat{\chi}(x)+\hat{S}(x)e^{\hat{T}(x)}
\end{eqnarray}
\noindent
where $\chi$ and $S$ depend only on momentum space variables, whose specific form can be read off from (\ref{ALGEBRA11}).  Note that the following relations hold
\begin{eqnarray}
\label{SHOWIT1}
\bigl[\hat{\chi}(x),e^{\hat{T}(y)}\bigr]=-\bigl[e^{\hat{T}(y)},\hat{\chi}(x)\bigr]=\Bigl({{\partial\chi} \over {\partial\Pi}}\Bigr)_xe^{\hat{T}(y)}\delta^{(3)}(x,y).
\end{eqnarray}
\noindent
This is a consequence of (\ref{THEFOLLOWINGONE}), where $\Pi$ is the only variable with nonvanishing relations with $T$.  We are now ready to proceed with the fourth term of (\ref{ALGEBRA13}), which is given by
\begin{eqnarray}
\label{NEWER}
\int_{\Sigma}d^3x\int_{\Sigma}d^3y\hat{W}(x,y)\Bigl[\hat{\chi}(x)+\hat{S}(x)e^{\hat{T}(x)},\hat{\chi}(y)+\hat{S}(y)e^{\hat{T}(y)}\Bigr]
\end{eqnarray}
\noindent
where we have used the definitions (\ref{SHOWIT}).  Expansion of (\ref{NEWER}) leads to the following four terms
\begin{eqnarray}
\label{SHOWIT2}
\int_{\Sigma}d^3x\int_{\Sigma}d^3y\hat{W}(x,y)\bigl[\hat{\chi}(x),\hat{\chi}(y)\bigr]
+\int_{\Sigma}d^3x\int_{\Sigma}d^3y\hat{W}(x,y)\hat{S}(x)\bigl[\hat{\chi}(x),e^{\hat{T}(y)}\bigr]\nonumber\\
+\int_{\Sigma}d^3x\int_{\Sigma}d^3y\hat{W}(x,y)\hat{S}(x)\bigl[e^{\hat{T}(x)},\hat{\chi}(y)\bigr]
+\int_{\Sigma}d^3x\int_{\Sigma}d^3y\hat{W}(x,y)\bigl[\hat{S}(x)e^{\hat{T}(x)},\hat{S}(y)e^{\hat{T}(y)}\bigr],
\end{eqnarray}
\noindent
which we will in turn analyse.  The first term of (\ref{SHOWIT2}) vanishes due to vanishing commutation relations between momentum space variables.  Using the results of (\ref{SHOWIT1}), the middle two terms of (\ref{SHOWIT2}) combine into
\begin{eqnarray}
\label{SHOWIT3}
\int_{\Sigma}d^3x\int_{\Sigma}d^3y\hat{W}(x,y)\Bigl[-\Bigl({{\partial\hat{\chi}} \over {\partial\hat{\Pi}}}\Bigr)_xe^{\hat{T}(y)}+\Bigl({{\partial\hat{\chi}} \over {\partial\hat{\Pi}}}\Bigr)_ye^{\hat{T}(x)}\Bigr]\delta^{(3)}(x,y)\nonumber\\
=\int_{\Sigma}d^3x\hat{W}(x,x)\Bigl[-\Bigl({{\partial\hat{\chi}} \over {\partial\hat{\Pi}}}\Bigr)_xe^{\hat{T}(x)}+\Bigl({{\partial\hat{\chi}} \over {\partial\hat{\Pi}}}\Bigr)_xe^{\hat{T}(x)}\Bigr]=0
\end{eqnarray}
\noindent
which vanishes after integration of the delta function, leaving remaining the fourth term of (\ref{SHOWIT2}).  Application of the identity (\ref{ALGEBRA10}) to this term yields
\begin{eqnarray}
\label{SHOWIT4}
\int_{\Sigma}d^3x\int_{\Sigma}d^3y\hat{W}(x,y)\bigl[\hat{S}(x)e^{\hat{T}(x)},\hat{S}(y)e^{\hat{T}(y)}\bigr]\nonumber\\
=\int_{\Sigma}d^3x\int_{\Sigma}d^3y\hat{W}(x,y)\hat{S}(x)\bigl[\hat{S}(x),e^{\hat{T}(y)}\bigr]e^{\hat{T}(x)}
+\int_{\Sigma}d^3x\int_{\Sigma}d^3y\hat{W}(x,y)\hat{S}(x)\bigl[e^{\hat{T}(x)},\hat{S}(y)\bigr]e^{\hat{T}(y)}\nonumber\\
+\int_{\Sigma}d^3x\int_{\Sigma}d^3y\hat{W}(x,y)\bigl[\hat{S}(x),\hat{S}(y)\bigr]e^{\hat{T}(x)}e^{\hat{T}(y)}
+\int_{\Sigma}d^3x\int_{\Sigma}d^3y\hat{W}(x,y)\hat{S}(x)\hat{S}(y)\bigl[e^{\hat{T}(x)},e^{\hat{T}(y)}\bigr].\nonumber\\
\end{eqnarray}
\noindent
The third term of (\ref{SHOWIT4}) vanishes due to vanishing commutation relations between momentum space variables, and the fourth term vanishes due to vanishing 
commutation relations between configuration space variables $T$.  Using (\ref{SHOWIT1}), the first and second term of (\ref{SHOWIT4}) combine into
\begin{eqnarray}
\label{SHOWIT5}
\int_{\Sigma}d^3x\int_{\Sigma}d^3y\hat{W}(x,y)\Bigl[-\hat{S}(y)\Bigl({{\partial\hat{S}} \over {\partial\hat{\Pi}}}\Bigr)_xe^{\hat{T}(x)}
+\hat{S}(x)\Bigl({{\partial\hat{S}} \over {\partial\hat{\Pi}}}\Bigr)_ye^{\hat{T}(y)}\Bigr]\delta^{(3)}(x,y)\nonumber\\
=\int_{\Sigma}d^3x\int_{\Sigma}d^3y\hat{W}(x,y)\Bigl[-S\Bigl({{\partial{S}} \over {\partial\Pi}}\Bigr)e^T+S\Bigl({{\partial{S}} \over {\partial\Pi}}\Bigr)e^T\Bigr]=0
\end{eqnarray}
\noindent 
which also vanishes.  We have shown that the third and fourth terms on the right hand side of (\ref{ALGEBRA13}) both vanish, which leaves us with the first and second terms.  Make the definitions
\begin{eqnarray}
\label{ALGEBRA14}
{{\delta\Phi(x)} \over {\delta\Pi_1(y)}}=Q^1(x)\delta^{(3)}(x,y);~~
{{\delta\Phi(x)} \over {\delta\Pi_2(y)}}=Q^2(x)\delta^{(3)}(x,y);~~
{{\delta\Phi(x)} \over {\delta\Pi(y)}}=Q^3(x)\delta^{(3)}(x,y),
\end{eqnarray}
\noindent
where $Q^i(x)$ are functions on $\Omega_{Kin}$ whose specific form will not be needed for what follows.  Likewise make the definitions
\begin{eqnarray}
\label{ALGEBRA15}
{{\delta\eta(x)} \over {\delta{X}(y)}}=\zeta^j_1(x)\delta^{(3)}(x,y)\Bigl({\partial \over {\partial{x}^j}}\Bigr);\nonumber\\
{{\delta\eta(x)} \over {\delta{Y}(y)}}=\zeta^j_2(x)\delta^{(3)}(x,y)\Bigl({\partial \over {\partial{x}^j}}\Bigr);\nonumber\\
{{\delta\eta(x)} \over {\delta{T}(y)}}=\zeta^j_3(x)\delta^{(3)}(x,y)\Bigl({\partial \over {\partial{x}^j}}\Bigr).
\end{eqnarray}
\noindent
The notation in (\ref{ALGEBRA15}) signifies that the partial derivatives will act on all objects with $x$ dependence which multiply the terms that the derivatives originally came from.  Using (\ref{ALGEBRA14}) and (\ref{ALGEBRA15}), we have the following operator relations
\begin{eqnarray}
\label{ALGEBRA16}
\bigl[\hat{\zeta}(x),\hat{\Phi}(y)\bigr]=\hat{\zeta}^j_I(x)\hat{Q}^I(y)\delta^{(3)}(x,y)\Bigl({\partial \over {\partial{x}^j}}\Bigr);\nonumber\\
\bigl[\hat{\Phi}(x),\hat{\zeta}(y)\bigr]=-\hat{\zeta}^j_I(y)\hat{Q}^I(x)\delta^{(3)}(x,y)\Bigl({\partial \over {\partial{y}^j}}\Bigr).
\end{eqnarray}
\noindent
Hence it is apparent from (\ref{ALGEBRA16}) that ${\partial \over {\partial{x}^i}}$ acts on objects containing $x$ dependence, and ${\partial \over {\partial{y}^i}}$ acts on objects containing $y$ dependence.  So continuing 
from (\ref{ALGEBRA13}) and using (\ref{ALGEBRA16}), we have
\begin{eqnarray}
\label{ALGEBRA17}
\bigl[\hat{H}[M],\hat{H}[N]\bigr]=\nonumber\\
\int_{\Sigma}d^3x\int_{\Sigma}d^3yM(x)N(y)\Biggl[\hat{\eta}(y)\hat{\zeta}^j_I(x)\hat{Q}^J(y)\Bigl({\partial \over {\partial{x}^j}}\Bigr)\hat{\Phi}(x)\nonumber\\
-\hat{\eta}(x)\hat{\zeta}^j_I(y)\hat{Q}^J(x)\Bigl({\partial \over {\partial{x}^j}}\Bigr)\hat{\Phi}(y)\Bigr]\delta^{(3)}(x,y)\nonumber\\
=\int_{\Sigma}d^3x\int_{\Sigma}d^3y\biggl[N(y)\hat{\eta}(y){\partial \over {\partial{x}^j}}(M(x)\hat{\zeta}^j_I(x)\hat{Q}^I(y)\hat{\Phi}(x))\nonumber\\
-M(x)\hat{\zeta}(x){\partial \over {\partial{y}^j}}(N(y)\hat{\zeta}^j_I(y)\hat{Q}^I(x)\hat{\Phi}(y))\biggr]\delta^{(3)}(x,y).
\end{eqnarray}
\noindent
Integration with respect to $y$ collapses the delta function, which yields
\begin{eqnarray}
\label{ALGEBRA18}
\int_{\Sigma}d^3x\Bigl[N\hat{\zeta}{\partial \over {\partial{x}^j}}(M\hat{\zeta}^j_I\hat{Q}^I\hat{\Phi})
-M\hat{\zeta}{\partial \over {\partial{x}^j}}(N\hat{\zeta}^j_I\hat{Q}^I\hat{\Phi})\Bigr]\nonumber\\
=\int_{\Sigma}d^3x\bigl(N\partial_iM-M\partial_iN\bigr)\hat{\eta}\hat{\zeta}^j_I\hat{Q}^I\hat{\Phi},
\end{eqnarray}
\noindent
whence the operator $\hat{\Phi}$ appears to the right.  Since $\hat{\Phi}$ is proportional to the Hamiltonian constraint it follows that 
\begin{eqnarray}
\label{ALGEBRA19}
\bigl[\hat{H}[M],\hat{H}[N]\bigr]\bigl\vert\psi\bigr>=\hat{H}[M,N]\bigl\vert\psi\bigr>,
\end{eqnarray}
\noindent
namely that the commutator of two Hamiltonian constraints is a Hamiltonian constraint with the constraint appearing to the right.  The quantum algebra of the Hamitonian constraint closes with structure functions, and it closes in direct analogy to its classical counterpart in \cite{EYOITA} when one makes the identification $q^I\eta^j_I\rightarrow\hat{\eta}\hat{\zeta}^j_I\hat{Q}^I$.  Moreover, the algebra closes with the proper ordering taken into account with the Hamiltonian constraint operator to the right.  For these reasons we conclude that the quantum constraints algebra is Dirac consistent and is free of anomalies.

\section{Conclusion}

The main result of this paper has been to verify the closure of the quantum constraints algebra for a theory of `reduced' gravity introduced in \cite{EYOITA}.  Future directions of research will be to investigate the Hibert space structure of the resulting theory.

\end{document}